\documentclass[aps,superscriptaddress,,twocolumn,showpacs]{revtex4-1}

\pdfoutput=1

\usepackage{color}

\usepackage{amsmath, amsthm, amsfonts}    % need for subequations
\usepackage{amssymb}
\usepackage{mathptmx} 
\usepackage{dsfont}
\usepackage{tikz}

\usepackage[unicode=true,pdfusetitle,
 bookmarks=true,bookmarksnumbered=false,bookmarksopen=false,
 breaklinks=true,pdfborder={0 0 0},backref=false,colorlinks=true,citecolor=blue]{hyperref}

\usepackage{graphicx}   % need for figures
\usepackage[caption=false]{subfig}       % use for side-by-side figures and   to have captions justified and small letter size

\usepackage{bm}            % bold math
\usepackage[normalem]{ulem}  % use for underlining

\newlength\imagewidth
\newlength\imagescale

\newcommand\numberthis{\addtocounter{equation}{1}\tag{\theequation}}
\def\be{\begin{eqnarray}}
\def\ee{\end{eqnarray}}

\definecolor{JOT-color}{named}{blue}
\definecolor{CSF-color}{named}{orange}

\begin{document}

%\title{Revisiting the so-called Second Kerker condition}

\title{Scattering asymmetry and second Kerker condition}

%\title{Does the so-called Second Kerker condition minimize the forward scattering?}
%
%\title{Dismantling the so-called Second Kerker condition}
%
%\title{New insights of the so-called Second Kerker condition}

\author{Jorge Olmos-Trigo}
\email{jolmostrigo@gmail.com}
\affiliation{Donostia International Physics Center (DIPC),  20018 Donostia-San Sebasti\'{a}n, Basque Country, Spain}

\author{Diego R. Abujetas}
\affiliation{Donostia International Physics Center (DIPC),  20018 Donostia-San Sebasti\'{a}n, Basque Country, Spain}
\affiliation{Instituto de Estructura de la Materia (IEM-CSIC), Consejo Superior de Investigaciones Cient\'{\i}ficas, Serrano 121, 28006 Madrid, Spain}

\author{Cristina Sanz-Fern\'andez}
\affiliation{Centro de F\'{i}sica de Materiales (CFM-MPC), Centro Mixto CSIC-UPV/EHU,  20018 Donostia-San Sebasti\'{a}n, Basque Country, Spain}

\author{Jos\'e A. S\'anchez-Gil}
\affiliation{Instituto de Estructura de la Materia (IEM-CSIC), Consejo Superior de Investigaciones Cient\'{\i}ficas, Serrano 121, 28006 Madrid, Spain}

\author{Juan Jos\'e S\'aenz}
\affiliation{Donostia International Physics Center (DIPC),  20018 Donostia-San Sebasti\'{a}n, Basque Country, Spain}
\affiliation{IKERBASQUE, Basque Foundation for Science, 48013 Bilbao, Basque Country, Spain}

\begin{abstract}
The nearly zero optical forward scattering and anti-dual conditions are usually associated to  the so-called second Kerker condition, at which the electric and magnetic responses are phase-shifted by $\pi$. However, as we  show,  this condition is insufficient  to both give rise to the nearly zero optical forward scattering and the anti-duality symmetry,  in striking contrast to the actual view of the problem. Interestingly, we demonstrate that near the  electric and magnetic dipolar resonances, the energy radiation pattern in the far-field limit resembles to the one arising from the first Kerker condition, with nearly-zero  backscattering.
\end{abstract}

\maketitle    
%\setstretch{1}
The conditions of perfect zero light  scattering in backward and forward directions were brought to the physical scene by Kerker, Wang and Giles  \citep{kerker1983electromagnetic} by assuming unusual magneto-dielctric  spheres that present both electric permittivity, $\epsilon$, and magnetic  permeability, $\mu$. 
In that work, it was proved that when $\epsilon = \mu$, the backscattered radiation from the sphere is identical to zero. In contrast, when $\epsilon = - \mu$, the forward scattering was shown to be reduced to zero.

The study of these two optical responses, known  as the first and second  Kerker conditions,  has attracted a great interest during the last decades \citep{mehta2006experimental, garcia2008exception,garcia2008light, laskar2008light, garcia2011directionality}. Recently, it has been shown that  high refractive index (HRI) materials \citep{garcia2011strong,kuznetsov2012magnetic, alaee2015generalized}, that present a null magnetic response ($\mu = 1$), can give rise to the aforementioned Kerker conditions. Indeed, when the electric and magnetic Mie coefficients, 
\begin{align}\label{Mie_tot}
a_l = i \sin \alpha_l e^{-i \alpha_{l}}, && b_l = i \sin \beta_l e^{-i \beta_{l}}, 
\end{align}
where  $\alpha_l$ and $\beta_{l}$ are the so-called electric and magnetic scattering phase shifts, are identical, i.e. $a_l = b_l \Longleftrightarrow \alpha_l = \beta_l$, the zero optical backscattering condition is  fulfilled \citep{geffrin2012magnetic, person2013demonstration,fu2013directional,
liu2013scattering,liu2014ultra,wang2017enhanced}. This physical phenomenon is associated  to the maximization of the asymmetry parameter  in the electromagnetic dipolar approximation \citep{olmos2019role}. Interestingly, under {illumination with a well-defined helicity beam}, this condition is as well linked with the conservation of the dual symmetry  \citep{ fernandez2013electromagnetic,olmos2019enhanced}.
In the same direction, Kerker conditions were found in transverse direction for directional coupling \citep{neugebauer2016polarization,picardi2018janus,shamkhi2019transverse}, Angstrom localization \citep{bag2018transverse} and control  of waveguide modes \citep{nechayev2019huygens}.

However, the condition $a_l = - b_l$, that would give rise to the  zero optical forward scattering, is {inhibited} by the optical theorem \citep{alu2010does}.
This special optical response would {ideally} return the minimum analytical value of the asymmetry parameter in the electric and magnetic dipolar regime or equivalently, would lead to the perfect antiduality \citep{zambrana2013dual, zambrana2013duality}. To ensure the energy conservation,  the idea of the anti-crossing of the electric and magnetic phase shifts, $\alpha_l = -\beta_l $, which leads to $\Re \{a_l \} = \Re \{b_l \}$ and $\Im \{a_l \} = - \Im \{b_l \}$, according to Eq.\ref{Mie_tot},  was suggested as the new second Kerker condition \citep{nieto2011angle}. This  optical response has been interpreted as the optimized condition that may  give rise to a negative asymmetry parameter \citep{gomez2012negative,
liu2018polarization,wang2018achieving} and reduce the scattering  in the forward direction 
\citep{pors2015unidirectional,
schmidt2015isotropically,naraghi2015directional,
decker2016resonant,genevet2017recent, kivshar2017meta}. This intriguing effect has been discussed as well in the contexts of optical forces \citep{nieto2010optical,aunon2014optical, nieto2015optical, gao2017optical, lank2018directional} and in the novel concept of the anapole modes \citep{luk2017hybrid,luk2017suppression,
liu2018generalized}.
This represents 
the  status of the so-called  second Kerker condition.

In this Letter, we  demonstrate that the current second Kerker condition, which is based on the anti-crossing of the electric and magnetic scattering phase-shifts, $\alpha_1 = -\beta_1$, does not lead to the (nearly) zero optical forward  condition for strong scattering regimes.  From the differential scattering cross section, which encodes the far-field radiation pattern via the asymmetry parameter $g$, we discuss the  origin of this misunderstanding. Our results reveal that the second Kerker condition straightforwardly  returns the minimum $g$-parameter for a fixed scattering cross section. However, it is not always negative, fact that inhibits the induction of the (nearly) anti-duality symmetry.  As a result, the differential scattering cross section or, in other words, the re-distribution of the energy in the far-field limit, ranges all possible scattering angle diagrams. Particularly, we show that near the electric and magnetic dipolar  resonances, the differential scattering cross section resembles  the one given by the first Kerker condition, where there is no net radiation in the backscattering direction, in opposition to the physical insight given until the date.

\begin{figure}[]
\includegraphics[width= \columnwidth]{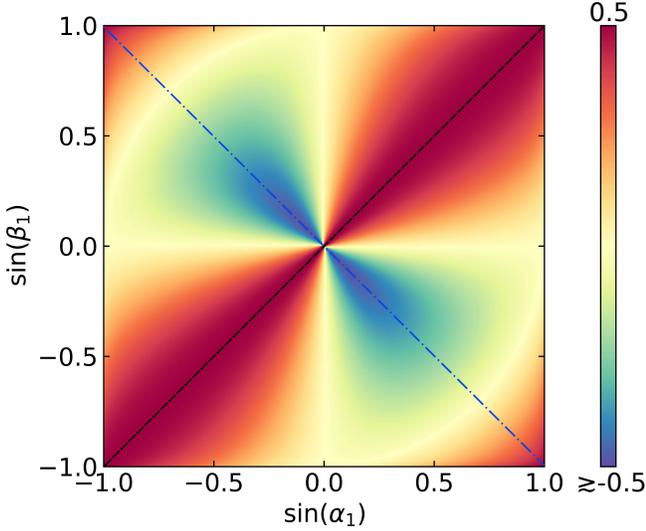}
\captionsetup{justification= raggedright}
\caption{Asymmetry parameter, $g$, as a function of the dipolar electric and magnetic  scattering phase-shifts, $\alpha_1$ and $\beta_1$, respectively.  The first Kerker condition (black dashed  line), given by $\alpha_1 = \beta_1$,  gives rise to the maximum  value of the $g$-parameter, regardless of the total scattering cross section. As seen in the attached scale, the second Kerker condition (blue dash-dotted  line), given by $\alpha_1 = -\beta_1$, minimizes the $g$-parameter for a fixed scattering cross section, $\sigma_{\rm{s}}$ (see Eq.~\eqref{crossdip}). The minimum (analytical) value, given by
$g = −1/2$, cannot be reached due to causality.}\label{1}
\end{figure}

The electromagnetic fields  scattered by high refractive index (HRI) dielectric nanoparticles present curious properties arising from the interference
between the electric and magnetic multipoles. Most of them  are embedded in the asymmetry parameter $g$, calculated from the differential scattering cross section, as
\begin{equation}
g = \langle \cos \theta \rangle = \frac{\int \frac{d \sigma_{\rm{s}}}{d \Omega} \cos \theta \; d \Omega}{\int \frac{d \sigma_{\rm{s}}}{d \Omega} \; d \Omega} = \frac{\int \frac{d \sigma_{\rm{s}}}{d \Omega} \cos \theta \; d \Omega}{\sigma_{\rm{s}}},
\end{equation}
where
 $\sigma_{\rm{s}}$ is the scattering cross section \citep{bohren2008absorption}.

Let us now consider the scattering in a spectral range such that the optical response
can be described by the first dipolar ($l=1$) electric and magnetic Mie coefficients of the particle,
\begin{align}\label{Mie}
a_1 = i \sin \alpha_1 e^{-i \alpha_{1}}, && b_1 = i \sin \beta_1 e^{-i \beta_{1}}.
\end{align}
This dipolar response can be achieved either by using a plane wave impinging on a small magnetoelectric particle, or illuminating a  larger sphere by a (sectoral)  dipolar beam \citep{olmos2019sectoral}, that, by construction, discards a higher multipolar response \citep{zaza2019size,trigo2019helicity}. In both cases, it is easy to show  that the differential scattering cross section is given by \citep{olmos2019enhanced}
\begin{equation} \label{dcross}
\frac{d \sigma_{\rm{s}} \left( \theta \right)}{d \Omega}  = \frac{3}{8 \pi} \sigma_{\rm{s}}\left( \frac{1 + \cos^2 \theta}{2} + 2g \cos \theta \right),
\end{equation}
where
\begin{equation}\label{g}
g = \frac{\Re \{ a_1 b^*_1 \}}{|a_1|^2 + |b_1|^2 } =   \frac{\sin \alpha_1 \sin \beta_1 \cos \left( \alpha_1  - \beta_1 \right) }{\sin^2 \alpha_1 + \sin^2 \beta_1 },
\end{equation}
is the asymmetry parameter \citep{bohren2008absorption} and 
\begin{equation}\label{crossdip}
\sigma_{\rm{s}}   = \frac{6 \pi}{k^2}  \left(|a_1|^2 + |b_1|^2 \right)=  \frac{6 \pi}{k^2}  \left( \sin^2 \alpha_1 + \sin^2 \beta_1   \right)
\end{equation}
is the scattering cross section \citep{hulst1957light}, both in the electric and magnetic regime .

\begin{figure}[]
\includegraphics[width=0.892 \columnwidth]{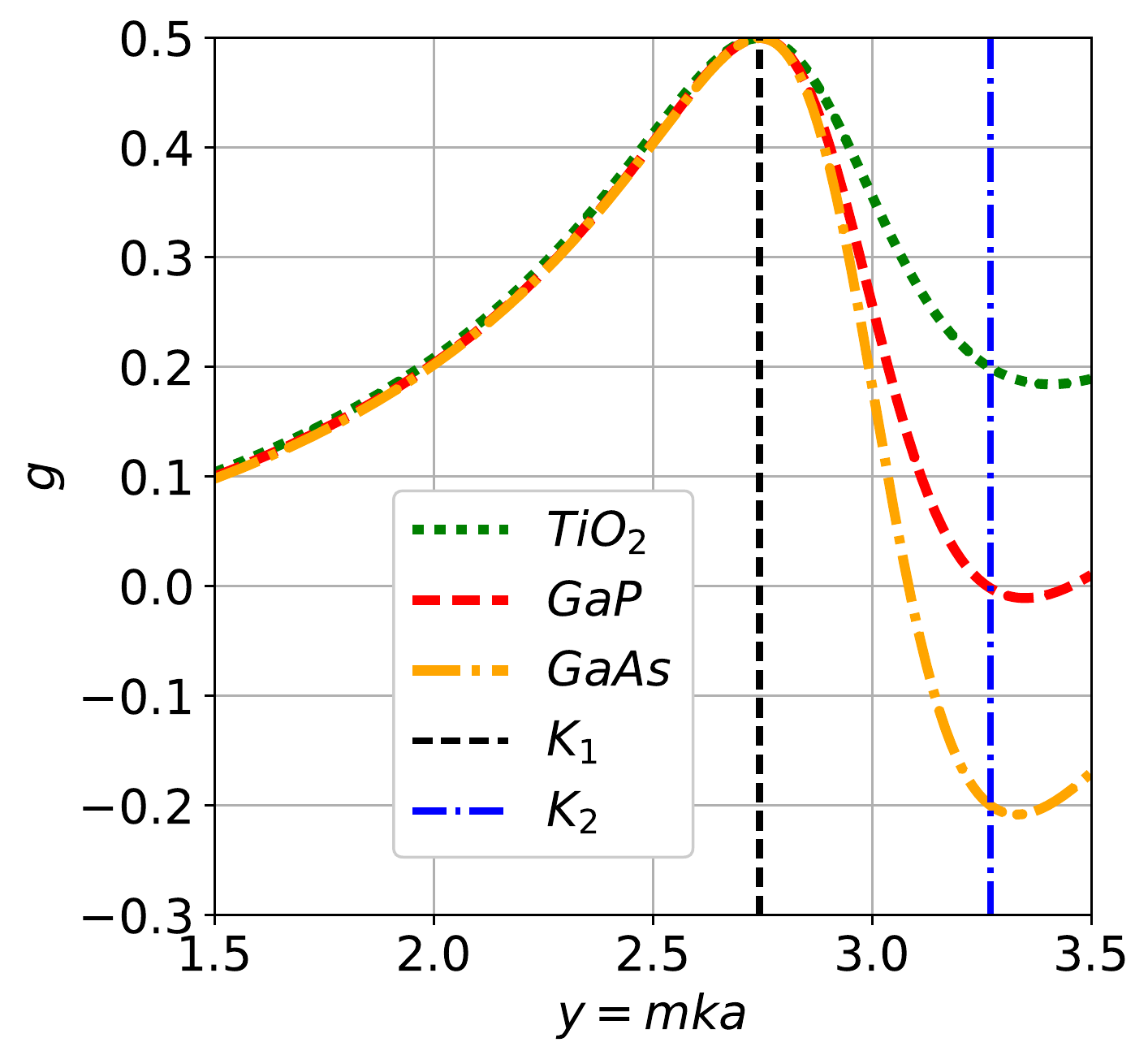}
\captionsetup{justification= raggedright}
\caption{Asymmetry parameter, $g$, as a function of the $y = mka$ size parameter, where $m$ is the relative  refractive index, $k$ is the wavevector in the medium and $a$ is the radius of the target. The black and blue dashed and dash-dotted lines specifies the first and second Kerker conditions, respectively. At the second Kerker conditon ($K_2$), the $g$-parameter is negative for the Gallium Arsenide-like sphere (GaAs), roughly zero for the Gallium phosphide-like sphere (GaP),  case while it is positive for the Titanium Oxide-like sphere (TiO$_2$). }\label{Tip}
\end{figure}

\begin{figure*}[]
\includegraphics[width= \textwidth]{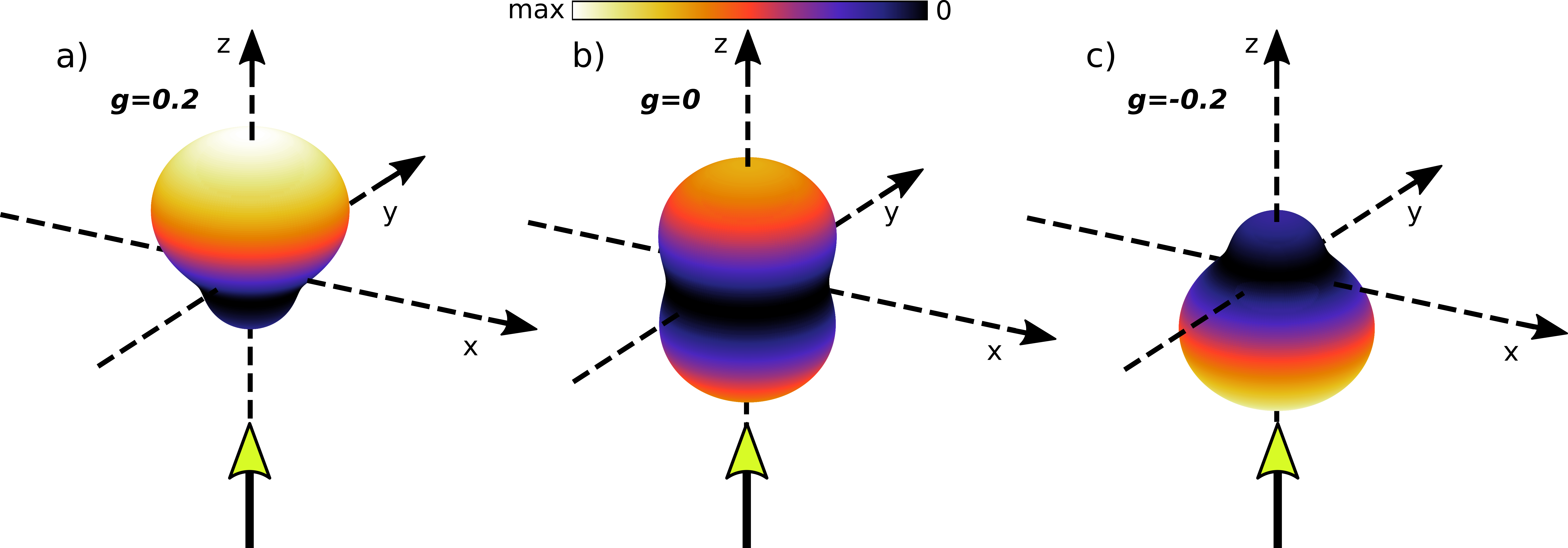}
\captionsetup{justification= raggedright}
\caption{(Integral-normalized) differential scattering cross section, $({{d \sigma_{\rm{s}} \left( \theta \right)} / {d \Omega}}) / {\sigma^{K_2}_{\rm{s}}}$,  for different optical responses at the second Kerker conditon, i.e $g = 0.2$ (TiO$_2$-like sphere), $g = 0$ (GaP-like sphere), $g = -0.2$ (GaAs-like sphere), according to Fig.~ \ref{Tip}. The incident light is illustrated by a vertical yellow arrow incising from backscattering.}\label{peras}
\end{figure*}

It is worth mentioning that the asymmetry parameter is equivalent, in the electric and magnetic dipole regime, to measure the degree of circular polarization (DoCP) at the perpendicular direction to the incoming wave, when excited by a well-defined helicity beam $\sigma = \pm 1 $ \citep{olmos2019asymmetry}. As a consequence, since the DoCP is the ratio of the $V$ and $I$  Stokes parameters, the asymmetry parameter is a straightforward measurable quantity  \citep{crichton2000measurable}.

From Eqs.~\eqref{dcross} and \eqref{g}, it  is trivial to see that, at the first Kerker condition, i.e., $g = 1/2 \Longleftrightarrow \alpha_1 = \beta_1$, the differential scattering cross section vanishes at backscattering {, i.e., $d \sigma_{\rm{s}} \left( \pi \right) / d \Omega = 0$,} regardless of the total scattering cross section, $\sigma_{\rm{s}}$. Interestingly, it can be shown that this does not  depend on the incident polarization \citep{fernandez2012helicity}. On the other hand, the optical response given by $ a_1 =  -b_1$ would lead to the minimum analytical value of the asymmetry parameter in the dipolar electric and magnetic regime, $g = -0.5$ \citep{arruda2016electromagnetic,arruda2017electromagnetic,
ali2018optimizing}. This in turn,  would give  rise to the perfect zero forward scattering condition, $d \sigma_{\rm{s}} \left( 0 \right) / d \Omega = 0$, according to Eq.~\eqref{dcross}. However, this condition is prohibited due to the optical theorem  \citep{nieto2015opticaltheorem,alu2010does}, which guarantees that the real part of the Mie coefficients must be positive-defined.
On these bases,  a new energetically viable second Kerker condition was brought into the physical scene by Nieto-Vesperinas \textit{et al.:} $\alpha_1 = -\beta_1$  \citep{nieto2011angle}. So far,   this was thought to be the optimized condition that  gives rise to a negative asymmetry parameter, which in turn might reduce the scattered light in the forward direction \citep{garcia2015all}.
Surprisingly, it is straightforward to notice via Eq.~\eqref{g} that the  second Kerker condition  does not (generally) lead to a negative asymmetry parameter,
\begin{align}\label{key}
\alpha_1 = -\beta_1 && \Longrightarrow &&
g = \frac{1}{2} \left[ \frac{k^2}{6 \pi} \sigma^{K_2}_{\rm{s}} - 1 \right],
\end{align}
where $\sigma^{K_2}_{\rm{s}} = (12 \pi / k^2) \sin ^2 \alpha_1$, according to Eq.~\eqref{crossdip}.

This is the first  important result of the present work. 
As it can be seen in   Eqs.~\eqref{crossdip} and \eqref{key}, just a relatively weak scattering  leads to negative values of the $g$-parameter.  The threshold, i.e., $g = 0$, is given by $\pm \alpha_1 =  \mp  \beta_1 = \pi/4$. Interestingly, this value corresponds to the scattering cross section that arises from a  pure dipolar electric (or magnetic) resonant particle, $\sigma^{\rm{res}}_{\rm{s}} = 6 \pi / k^2 $.

This phenomenology can be inferred in Fig.~\ref{1}, where the $g$-parameter is illustrated as a function of the dipolar electric and magnetic scattering phase-shifts. Notice that  Fig.~\ref{1} covers all possible optical responses in the electromagnetic dipolar regime as they run over all possible values of the first electric and magnetic Mie coefficients,  according to Eq.~\eqref{Mie}. It is important to recall that the Mie coefficients can be generalized for any geometry \citep{abujetas2018generalized}. As predicted, the first Kerker condition  ($\alpha_1 =  \beta_1$)  gives rise to the maximum value of the asymmetry parameter, $g =  0.5 $. In addition, this is completely independent of the  scattering cross section, $\sigma_{\rm{s}}$, which corresponds to  circles in the figure, according to Eq.~\eqref{crossdip}. Interestingly, it can be inferred that the so-called second Kerker condition, $\alpha_1 =  -\beta_1$, minimizes the asymmetry parameter for a fixed scattering cross section. However, this is not sufficient to state that this condition always leads to negative values of $g$. As an immediate consequence, the (nearly) anti-duality symmetry is not generally satisfied  \citep{zambrana2013dual, olmos2019asymmetry}.
As an illustrative example that confirms the previous statement, the asymmetry parameter, is considered for three spheres of different materials, Titanium Oxide-like (TiO$_2$), Gallium Phosphide-like (GaP) and Gallium Arsenide-like (GaAs),  in air, as it can be seen in Fig.~\ref{Tip}. Refractive index data for these materials were taken from \citep{aspnes1983dielectric}. This
figure shows that, at the first Kerker condition, $\alpha_1 =  \beta_1$, the
maximum value of the asymmetry is always reached (dashed vertical black line), leading to the zero backscattering condition. On the other hand,   at the second Kerker condition, $\alpha_1 = - \beta_1$, which corresponds to the  dash-dotted vertical blue line, the asymmetry parameter is not always negative. In fact, the asymmetry parameter is  positive for the TiO$_2$-like sphere, almost  zero for the GaP-like sphere case while it becomes negative for the GaAs-like sphere. {According to Eq.~\eqref{key}, this change of sign  depends strongly on the strength of the scattering cross section.}

Let us now analyse the relevance of the second Kerker condition in the (nearly) zero optical forward scattering condition, where
\begin{align}\label{cris}
{\frac{d \sigma_{\rm{s}} \left( \theta \right)}{d \Omega}} = \frac{3}{8 \pi} {\sigma^{K_2}_{\rm{s}}}\left( \frac{1 + \cos^2 \theta}{2} + \left[ \frac{k^2}{6 \pi}{\sigma^{K_2}_{\rm{s}}} - 1 \right] \cos \theta \right).
\end{align}
At the second Kerker condition, $\alpha_1 =  -\beta_1$, the scattering cross section, ${\sigma^{K_2}_{\rm{s}}}$, governs the far-field pattern of the differential scattering cross section. As a consequence, the second Kerker condition is not sufficient (not even necessary) to obtain the typical pear-like structure, in striking contrast to previous analysis \citep{gomez2011electric,garcia2013sensing}.
In fact, it is clear that near the electric and magnetic dipole resonances, in which $g \lesssim 0.5 \Longleftrightarrow \sigma^{K_2}_{\rm{s}} \lesssim 12 \pi / k^2$,  the radiation pattern of the  differential scattering cross section reminds to the one arising when the first Kerker condition is satisfied. In this particular case, there is no net radiation at backscattering. On the other hand, when $\sigma^{K_2}_{\rm{s}} = \sigma^{\rm{res}}_{\rm{s}} = 6 \pi / k^2$, condition that leads to $g = 0$, according to Eq.~\eqref{key}, the differential scattering cross section is identical to the one given by a pure electric (or magnetic) dipole. Finally, when $\sigma^{K_2}_{\rm{s}} < \sigma^{\rm{res}}_{\rm{s}}$, which implies a negative asymmetry parameter,   the target scatters mostly in the forward direction.

This behaviour is further confirmed in Fig.~\ref{peras}, where the far field radiation pattern of the  (integral normalized) differential scattering cross section is considered, at the second Kerker condition, for the materials illustrated in Fig.~\ref{Tip}:  Ti0$_2$-like, AsP-like and AsGa-like spheres. As  expected, when $g>0$, the scattering pattern almost entirely lies  in the forward direction, according to Fig.~\ref{peras}a . This  corresponds to the energy radiation pattern in the far field limit  of the Ti0$_2$-like sphere at the second Kerker condition, where $g = 0.2$, (see dotted
green line Fig.~\ref{Tip}). At $g = 0$, which arises from the GaP-like sphere at the second Kerker condition, according to the
dashed red line in Fig.~\ref{Tip},    the energy radiation pattern in the far field limit  is symmetrical, as it can be seen in Fig.~\ref{peras}b. This  is identical to the one that  arises from  a pure electric (or magnetic) dipole.  Finally,  when $ g < 0$, the target preferentially scatters in the backward direction, as it can be inferred in Fig.~\ref{peras}c.  This phenomenon corresponds to the energy radiation pattern in the far field limit of the GaAs-like sphere at the second Kerker condition, where $g = -0.2$, according to the dash-dotted yellow
line of  Fig.~\ref{Tip}.  This last case  corresponds to what was previously assumed as the natural implication of the second Kerker condition \citep{geffrin2012magnetic}. From our analysis it is straightforward to derive that both the  nearly anti-duality symmetry and the almost zero optical forward scattering condition are only achievable  through a negative $g$-parameter.

\begin{figure}[]
\includegraphics[width= 0.975 \columnwidth]{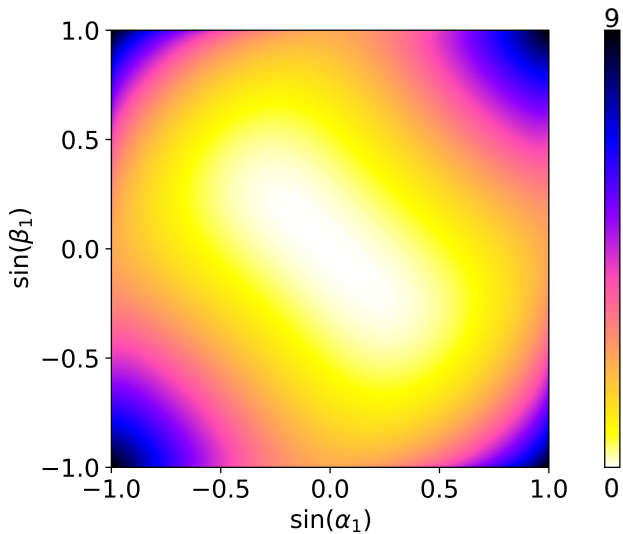}
\captionsetup{justification= raggedright}
\caption{Dimensionless differential scattering cross section, $k^2 {d {\sigma}_{\rm{s}}} / {d \Omega}$, evaluated at the forward direction, $\theta = 0$. At the second Kerker condition, $ \alpha_1 = - \beta_1$, near the electric and magnetic resonances, the dimensionless differential scattering cross section  is maximized. On the other hand, when the scattering cross section is relatively small, it reaches its minimum value.  }\label{3}
\end{figure}

In order to get deeper physical insight into the relevance of the second Kerker condition,   it is interesting to derive the explicit expression of the differential scattering cross section at the forward direction. Under these conditions, 
\begin{equation*}\numberthis \label{palante}
\begin{aligned}[c]
 \alpha_1 &=-\beta_1,\\
\theta&=0, 
\end{aligned}
\qquad\Longrightarrow\qquad
\begin{aligned}[c]
\frac{d {\sigma}_{\rm{s}}}{d \Omega}  &=  \frac{k^2}{16 \pi}   \left( \sigma^{K_2}_{\rm{s}} \right) ^2.
\end{aligned}
\end{equation*}
Equation~\eqref{palante} shows that at the second Kerker condition, in the forward direction, the differential scattering cross section scales quadratically with the scattering cross section. As a result, for a relatively large scattering cross section, the target may scatter preferentially in the forward direction, in striking contrast to the actual view of the problem. This can be seen in Fig.~\ref{3}, where the (dimensionless) differential scattering cross section, $k^2 {d {\sigma}_{\rm{s}}} / {d \Omega}$ , evaluated at the forward direction, $\theta = 0$, is illustrated.
As could be expected at the second Kerker condition, $\alpha_1 = - \beta_1$, the scattering in the forward direction is not minimized. In fact, near the electric and magnetic dipolar resonances, where $ \pm \alpha_1, \mp \beta_1 \approx \pi/2 \Longleftrightarrow g \lesssim 0.5$,  this is close to be  maximized. {Therefore, both Figs. \ref{1} and \ref{3} can be understood together as the actual implication of the second Kerker condition: Only when the scattering cross section is smaller than a pure resonant particle, $\sigma^{K_2}_{\rm{s}} < \sigma^{\rm{res}}_{\rm{s}}$,  a negative  $g$-parameter can be obtained. In this regime, the near zero optical forward scattering can be achieved.}

In conclusion,  we have shown that the current second Kerker condition can be derived as the optimal condition that  minimizes the asymmetry, in terms of the $g$-parameter, for a fixed scattering cross section. Interestingly, we have found that the second Kerker condition  does not necessarily give rise to a negative $g$-parameter, and hence the induction of anti-dual spheres. In fact, under this condition, we have demonstrated that the $g$-parameter ranges from positive to negative values, crossing $g = 0$ when the scattering cross section is identical to the one arising from a pure electric (or magnetic) resonant target.  As a direct consequence, we have demonstrated that the far-field scattering pattern of the  differential scattering cross section runs over all its possible polar diagrams. Notably, near the electric and magnetic dipole resonances, we have explicitly exposed that this resembles to the one given at the first Kerker condition, where there is no net radiation in the backscattering direction. This phenomena implies the opposite physical insight which was expected in previous works.

 In order to have a deeper insight, we have then confirmed this behavior by showing that the energy radiation  pattern, or in other words, the differential scattering cross section, under the second Kerker condition, scales quadratically with the scattering cross section. As a direct consequence, for strong scattering regimes, the target preferably scatters in the forward direction, in striking contrast with the current understanding. We do believe that our straightforward but fundamental analysis provides new insight in the study of light scattering from nanostructures which can be relevant in understanding more complex, multiple scattering processes in nano-structured samples and photonic devices. 

This research was supported by the Spanish Ministerio
de Econom\'{\i}a y Competitividad (MINECO, MICINN)
and European Regional Development Fund (ERDF)
through the grants LENSBEAM (Nos. FIS2015-69295-C3-2-P and FIS2015-69295-C3-3-P), NANOTOPO (No. FIS2017-91413-EXP),  FPU PhD Fellowship (No. FPU15/ 03566), MELODIA PGC2018-095777-B-C21, Project
FIS2017-82804-P
 and by the Basque Dep. de Educaci\'on
Project PI-2016-1-0041 and PhD Fellowship (PRE-
2018-2-0252).

\bibliographystyle{apsrev4-1}
\bibliography{New_era_18_06_2019.2}
\end{document}